\documentclass[runningheads,a4paper]{llncs}

\usepackage{amssymb}
\usepackage{graphicx}

\newcommand{\keywords}[1]{\par\addvspace\baselineskip
\noindent\keywordname\enspace\ignorespaces#1}

\usepackage{fancyhdr}
\begin{document}

\mainmatter  

\title{An Approach for Text Steganography\\Based on Markov Chains}

\titlerunning{An Approach for Text Steganography\\Based on Markov Chains}

%
%
\author{H. Hernan Moraldo}
\authorrunning{H. Hernan Moraldo}
\institute{}

\maketitle

\begin{abstract}
A text steganography method based on Markov chains is introduced, together with a reference implementation. This method allows for information hiding in texts that are automatically generated following a given Markov model. Other Markov - based systems of this kind rely on big simplifications of the language model to work, which produces less natural looking and more easily detectable texts. The method described here is designed to generate texts within a good approximation of the original language model provided.
\end{abstract}

\section{Introduction}

Steganography is the field that deals with the problem of sending a message from a sender A to a recipient B through a channel that can be read by a so-called Warden, in a way that the Warden doesn't suspect that the message is there.

Steganographic techniques exist for hiding messages in images, audio, videos, and other media. In particular, text steganography studies information hiding on texts. There are many techniques for this, as summarized on \cite{bennett} \cite{nechta}. One of the simplest steganographic methods for texts works by encoding a fixed amount of bits per word, using a table that maps words to codes, and vice versa. A disadvantage of this trivial technique is that the text will be obviously random at a syntactic level, as words are generated in a way that is independent of context.

There are other simple methods that store data in the text format, using spacing, capitalization, font or HTML tags. For example SNOW \cite{snow} hides information in tabs and spaces at the end of each line, that are usually not visible on text viewers. Also, some techniques start from a base text (the covertext), and modify it in some way: for example by switching words to near synonymous, or by changing sentences from their original grammatical structure to another one that preserves the meaning. The technique shown in \cite{topkara} hides information by modifying words in a way that resembles ortographical or typographical errors. There are other techniques that rely on translation \cite{grothoff}.

In other cases, texts are generated using a grammar model; this kind of system has the advantage of producing texts that make sense at a grammatical level, although not at a semantical level.

And there are techniques, like the one described on this paper, that are based on using Markov models to generate texts that encode some hidden message on them. Weihui Dai et al. \cite{weihui1} \cite{weihui2} explore a method for encoding data on this way; \cite{tentencouk} shows a simple implementation of a similar concept.

This article explores a specific method for using Markov chains for text steganography. How this method compares to other similar methods and how it works is explored further in the next sections. A reference implementation of the method described here is also included in the open source program MarkovTextStego \cite{moraldo}.

\section{Related Work}

Many methods for text steganography that are not based on Markov chains are known. An example is NiceText \cite{chapman}, which shows a way to encode ciphertext to text, that uses custom styles, Context Free Grammars and dictionaries.

The approach used in \cite{weihui1} \cite{weihui2} is based on Markov chains. When encoding, some data is provided as input, and the system generates a text as output using a given Markov chain. The stegotexts are generated in a way that simulates that they were generated by the Markov chain.

However, to avoid complex calculations, the Markov model is simplified by assuming that all probabilities from a given state to any other state are equal. This can change the quality of the texts generated by the Markov chain significantly. For example words like "the" and "naturally" are both potential starts of a phrase, but the former should be much more frequent than the latter; and this difference is not preserved by the simplification.

Other Markov - based models or similar models require of similar simplifications of the Markov chain, typically by making all outbound probabilities of each state equal (as in the previous example), or by replacing them by other ones, either explicitly or implicitly through the operation of the encoding algorithm \cite{tentencouk} \cite{siefkes} \cite{siefkes2}.

The method described here aims to be an answer to the question of whether it is possible to preserve the probabilities in the Markov models to higher levels of accuracy. The method is not optimally precise, but it generates texts that use a language model that is a good approximation of the provided Markov model.

\section{Markov Chain Models} \label{markovmodels}

A Markov chain is a model for a stochastic process. A sequence of random variables $X = (X_{1}, ..., X_{T})$ with values from a finite set $S$ is a Markov chain, if it has the Markov properties \cite{manning} \cite{russell}:

Limited Horizon property:

\begin{equation}
P(X_{t+1} = s_{k} | X_{1}, ..., X_{t}) = P(X_{t+1} = s_{k} | X_{t})
\end{equation}

Time Invariant property:

\begin{equation}
P(X_{t+1} = s_{k} | X_{t}) = P(X_{2} = s_{k} | X_{1})
\end{equation}

The first property means that the Markov chain doesn't have memory of any states, beyond the last one. The second property means that the conditional probabilities for all states do not depend on the position (time) on the sequence.

\begin{figure}
\centering
\includegraphics[height=4.7cm]{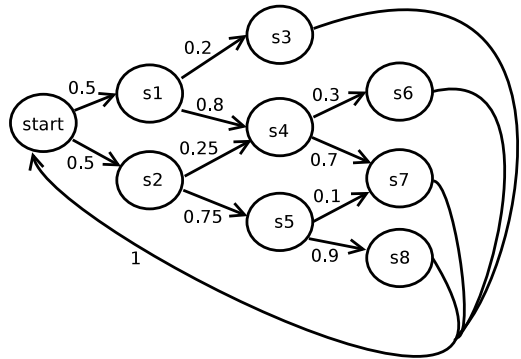}
\caption{Example Markov chain. In the context of text steganography, each state is a word. The special state "$start$" marks both the start and end of a sentence.}
\label{fig:examplechain}
\end{figure}

Diagrams like the one shown in Fig.~\ref{fig:examplechain} are frequently used to represent the transitions in Markov chains. All nodes in the graph represent states (elements of $S$), and any arrow from $s_{j}$ to $s_{k}$ with a value of $p$ means that $P(s_{k} | s_{j}) = p$. We call any state $s_{k}$ an outbound state of $s_{j}$, if there is an arrow from $s_{j}$ to $s_{k}$. For every $s_{j}$ that doesn't have an arrow to another $s_{k}$ state, $P(s_{k} | s_{j}) = 0$.

Markov chains and models are frequently used to model language \cite{manning}; when that's the case, states in the chain are used, for example, to represent words, characters, or n-grams. Also, Markov models are used in steganography (as described above), and in steganalysis \cite{sidorov} \cite{taskiran}.

A Markov language model may be useful to compute probabilities for phrases, from the n-gram probabilities. For example given the Markov chain shown in Fig.~\ref{fig:examplechain}, if the process were to start from "$start$" (symbol that we use both for start and end of a sentence), the probability of generating the text composed by the sequence of words or states $[s1, s4, s7]$ would be $0.28$.

These models can also be used to generate random texts. For this, a random source is used that can pick a next state $s_{k}$ with probability $P(s_{k} | s_{j})$, given the current state $s_{j}$. The algorithm for generating the random text starts by setting "$start$" to be the current state; then, in every iteration it uses the random source to pick the next word, which also becomes the new current state in the next iteration. To generate a single sentence, the process can be made to stop when the state "$start$" is reached.

Although Markov chains only have memory of a single previous state, every state can be a bigram, or an n-gram. This way, a Markov language model can have memory for more than a single word. Although this article only describes the steganographic method based on states that are single words, the reference implementation of the system \cite{moraldo} allows using both unigrams and bigrams as states, and it is possible to extend it to support n-grams with $n > 2$. Table~\ref{fig:tablefiles} compares the results of the encoding procedure when using unigrams and bigrams.

A Markov language model with states as single words can be computed from the frequencies of all bigrams, and all unigrams in a text:

\begin{equation}
P(w_{n} | w_{n-1}) = \frac{\displaystyle count(w_{n-1}, w_{n})}{\displaystyle count(w_{n-1})}
\end{equation}

where $count(a, b)$ is the number of occurrences of word $a$ followed immediately by word $b$ in the text, and $count(a)$ is the number of occurrences of word $a$.

As discussed above, some steganographic methods are based on these Markov language models. The language model is usually simplified in some way; for example \cite{weihui1} sets all $P(x | w)$ with $w$ fixed, to a fixed $k$.

In these models, once the simplification is done, the Markov chain is used to encode data into text; every word stores some fixed or variable amount of bytes, and every bigram in the generated text is required to have conditional probability $P(w_{n} | w_{n-1}) > 0$ in the Markov chain. A decoding algorithm that reverses the process, transforming the text into data, is also defined.

The approach shown in this article avoids much of the simplification in the probabilities of the Markov chain. Although there is still some precision loss in the model, for the most part, the proportions between the frequencies of different n-grams are preserved, specially for long texts.

\section{Fixed-size Steganography}

A main objective in this article is to describe two functions, $encode$, and $decode$, that are used to create a text out of a data input, and to get the original data out of an encoded text. In steganography literature, it would be said that $encode$ generates a stegotext out of the input plaintext, while $decode$ does the reverse process. The $encode$ function is not cryptographically secure; it assumes that its input is a plaintext, or some data that has already been encrypted using an independent system. In the latter case, $encode$'s input can be called ciphertext.

We require the encoding function to be invertible; that is, for every input data $d_{1}$ and $d_{2}$,  $encode(d_{1}) = encode(d_{2})$ only if $d_{1} = d_{2}$. Also $decode$ is the inverse of $encode$, so for every input $d$, $decode(encode(d)) = d$. The encoding function $encode$ is required to work on all the domain of data $d$; the required domain of $decode$ however needs only be the image of $encode$.

The $encode$ and $decode$ functions will be built out of simpler functions, for fixed-size encoding and decoding. These functions are $encode_{fixed}(data,$ $data size)$, and $decode_{fixed}(text, data size)$. Both the Markov chain and the starting symbol are actually required for these functions too, but they are left out for simplicity. Only when it is required for the purposes of the explanation, a third argument is added to both functions, for the start symbol: $encode_{fixed}(data,$ $data size,$ $start symbol)$, and $decode_{fixed}(text, data size, start symbol)$.

In this system, the size of $d$ in bits is known beforehand both for $encode_{fixed}$ and for $decode_{fixed}$. The requirements for both functions are weaken compared to those for their non-fixed counterparts; it is required that for every input data $d_{1}$ and $d_{2}$ such that $length(d_{1}) = length(d_{2})$,  $encode_{fixed}(d_{1}, length(d_{1})) = encode_{fixed}(d_{2}, length(d_{2}))$ only if $d_{1} = d_{2}$ (where $length(d)$ is the size of $d$ in bits). This weaker restriction means that the encoder may produce the same text for two different data inputs, only if they have different lengths, as can be seen in the examples in Table~\ref{fig:tableencodings}.

Also, $decode_{fixed}(z, length(z)) = d$ with $z = encode_{fixed}(d, length(d))$.

\subsection{Mapping of Probabilities to Ranges}\label{mappingprob}

A basic component for encoding and decoding is the function named $subranges$, that maps all outbound states from a given state, to subranges of a given range. These subranges are a partition of the original range.

\begin{equation} \label{subrangeseq}
subranges(mc, s, r) = [(s_{1}, r_{1}), ..., (s_{n}, r_{n})]
\end{equation}

where $mc$ is a Markov chain, $s$ is a state in $S$, and $r$ is a range of natural numbers $[a, b]$. The result is a list that must have some properties that are described below.

The behavior of this function is that it maps outbound states of a Markov chain to subranges of a given range, in a way that approximately matches the proportion between the sizes of the different subranges, to the proportion between the probabilities of the respective states. For example, if $mc$ is the chain in Fig.~\ref{fig:examplechain}, $s = start$, and $r = [0, 3]$, the expected result would be $[(s_{1}, [0, 1]), (s_{2}, [2, 3])]$. That is, because each outbound state has a 0.5 probability, it has to get half of the full range. If $s = s_{2}$, the expected result would be $[(s_{4}, [0, 0]),$ $(s_{5}, [1, 3])]$, where again the length of the subranges matches the proportion of their respective probabilities.

This partitioning method will be used in an iterative way, both for encoding and for decoding. Fig.~\ref{fig:exampleencoding} (in Section~\ref{encodingsection}) shows how this is done, although the details of the operation are described in the next sections.

The returned value for $subranges$ in Equation~\ref{subrangeseq} is a list of pairs $(s_{k}, r_{k})$, where $s_{k}$ is a state such that $P(s_{k} | s) > 0$, and $r_{k}$ is a subrange of $r$. The subranges of $r$ returned by $subranges$ are a partition of $r$.

A good implementation of the function generates a mapping between subranges $r_{k}$ and states $s_{k}$, such that the fraction of the total range length for each $r_{k}$ is approximately equal to the probability of the respective state $s_{k}$. That is:

\begin{equation} \label{conditionlen}
\frac{\displaystyle length(r_{k})}{\displaystyle length(r)} \approx P(s_{k} | s)
\end{equation}

Where the length of a range is defined to be $length([a, b]) = b - a + 1$.

The property in Equation~\ref{conditionlen} is not a strict requirement, as even without this condition the encoding function will still generate texts that are decoded correctly. However, only when this condition is held the texts that are generated will be approximately described by the original Markov language model.

\textbf{[Condition of Minimal Length]} The following condition is actually required for the steganographic system to work, however. Any time that there are at least two different states $s_{1}$ and $s_{2}$ such that $P(s_{1} | s) > 0$ and $P(s_{2} | s) > 0$ (that is, every time $s$ has at least two outbound states), it is required that $subranges$ returns a list with at least two elements.

This restriction is necessary for ensuring that both the encoder and decoder methods halt for all inputs. It might produce precision loss in many cases however, as in the following example: an $s$ state has two outbound states $s_{1}$ and $s_{2}$, with conditional probabilities $P(s_{1} | s) = 0.99$ and $P(s_{2} | s) = 0.01$, and the input range to process is $r = [0, 1]$.

In this case, it would seem that the best output would map $s_{1}$ to the full range: the returned value for this would be $[(s_{1}, [0, 1])]$. However this value doesn't hold the Condition of Minimal Length, as the list has a single element, despite $s$ having more than one outbound state.

Because of this, the only valid results for this example would be $[(s_{1}, [0, 0]),$ $(s_{2}, [1, 1])]$ and a symmetrical one (same ranges but switching states). As can be seen, these valid options are worse approximations to the input conditional probabilities, than just mapping $s_{1}$ to the full state; however the condition described disallows this better approximation.

The following three functions are used by the encoding and decoding methods.

As described, $subranges$ returns a list that maps ranges to states. The function $subrangeForState$ uses the list to return the subrange that is assigned to a given state:

\begin{eqnarray}
subrangeForState(mc, s_{k}, r, s_{l}) = r_{k}  \nonumber\\
\mbox{ from } [..., (s_{k}, r_{k}), ...] = subranges(mc, s_{k}, r)  \mbox{ such that } s_{k} = s_{l} 
\end{eqnarray}

The function $subrangeForNumber$ returns the subrange in the list that contains a given number:

\begin{eqnarray}
subrangeForNumber(mc, s_{k}, r, number) = \mbox{ subrange } r_{k} \nonumber\\
\mbox{ from } [..., (s_{k}, r_{k}), ...] = subranges(mc, s_{k},  r)  \mbox{ such that } number \in r_{k}
\end{eqnarray}

The function $stateForNumber$ returns the state that is assigned to the subrange returned by $subrangeForNumber$:

\begin{eqnarray}
stateForNumber(mc, s_{k}, r, number) = \mbox{ state } s_{k} \nonumber\\
 \mbox{ from } [..., (s_{k}, r_{k}), ...] = subranges(mc, s_{k}, r) \mbox{ such that } number \in r_{k} 
\end{eqnarray}

These functions will be used in the next sections.

\subsection{Encoding Fixed-size Data Using Markov Chains} \label{encodingsection}

The function $stateForNumber$, can also be seen as a function that encodes data to a single word. Given a Markov chain, a state, a range and a number (the input data), it finds the corresponding state or word in the chain for that number. Related to that, $subrangeForNumber$ also defined above, returns the subrange that corresponds to the word returned by $stateForNumber$.

Based on these two functions, a sequence of states $s_{t}$ and a sequence of ranges $r_{t}$ can be generated as described in the following two equations. These sequences are computed given a Markov chain $mc$, an initial state $s_{0}$ (typically "$start$"), an input data ($number$), and an initial range $r_{0}$ (typically $[0, 2^n - 1]$, where n is the length of the data to store):

\begin{equation}
s_{t} = stateForNumber(mc, s_{t-1}, r_{t-1}, number)
\end{equation}

\begin{equation}
r_{t} = subrangeForNumber(mc, s_{t-1}, r_{t-1}, number)
\end{equation}

Both sequences are defined to be finite (as we want to encode data to a finite sequence of words); the final element for both is $T$ such that $length(r_{T}) = 1$. This means that we stop encoding when the sequence of words describes a single number.

Finally, $encode_{fixed}(data, length(data)) = [s_{1}, ..., s_{T}]$.

This encoding process works by partitioning an input range in a way that matches the outbound states of a given state, and then selecting the outbound state whose subrange contains the number to encode. After this is done, the selected subrange and state are used as the input for the next iteration of the algorithm. When the process finishes, the encoded text is the sequence of states that the algorithm went through.

The $subranges$ function is restricted by the Condition of Minimal Length in Section~\ref{mappingprob} to always split a range in more than one subrange, whenever possible; therefore the iteration of this process will produce ranges that are smaller and smaller. (Even though it is possible that a Markov chain that is computed from a text contains states with only one outbound state, those will eventually lead to $start$, which will have more than one outbound state.) Also all subranges must contain at least one element, so the iterative generation of subranges converges to a subrange of length $1$.

Because the selected subrange length converges to $1$, the process has to finish, and when it finishes there is a subrange around a single number (the original input) and a list of states (words). For every input data, there is a final result.

For every number $d$ of size $n$, this final result can be seen as a path that points to $d$, as every state in the word sequence tells which subrange to choose from the partitions generated by $subranges$. Using this path intuition, it can be seen that if $d_{1}$ and $d_{2}$ are two different numbers of the same size $n$, their encoded texts are necessarily different, as they lead to different numbers. In the same way, the decoding system can find $d$ using the text as a path to the length $1$ subrange.

\begin{figure}
\centering
\includegraphics[height=3cm]{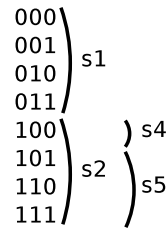}
\caption{Example encoding of $100$ into $[s_{2}, s_{4}]$. This and more examples can be seen in Table ~\ref{fig:tableencodings}.}
\label{fig:exampleencoding}
\end{figure}

Fig.~\ref{fig:exampleencoding} shows this partitioning process. The example in the figure uses the range $[0, 8]$, with the numbers encoded in binary. If we use the Markov chain shown in Fig.~\ref{fig:examplechain} and we start from $start$, in a first step the range has to be split in half, because the probabilities for the two states $s_{1}$ and $s_{2}$ are both $0.5$. The subrange assigned to $s_{2}$ can then be split in two other parts, now for the states $s_{4}$ and $s_{5}$, but the proportions are $0.25$ and $0.75$ in this case. This shows that if we were to encode the binary number $100$, with a fixed size $n = 3$ bits, we would get the text $[s_{2}, s_{4}]$. If we were trying to encode the binary number $111$, we would need to continue partitioning the range for $s{5}$, until there is only a single number in the last subrange.

Table ~\ref{fig:tableencodings} shows the output of $encode_{fixed}$ for a number of inputs. The reference implementation \cite{moraldo} was used, and the results may vary in other implementations, depending on specific details of the range partitioning algorithm. All examples use the Markov chain shown in Fig.~\ref{fig:examplechain}, with "$start$" as the starting state. In particular, it is possible to see that $100$ indeed encodes to $[s_{2}, s_{4}]$, as described above.

\begin{table}
\centering
\begin{tabular}{|c|c|c|}
	\hline
data & $n$ & encoded text \\
	\hline
0 & 1 & [s1] \\
1 & 1 & [s2] \\
00 & 2 & [s1, s3] \\
01 & 2 & [s1, s4] \\
10 & 2 & [s2, s4] \\
11 & 2 & [s2, s5] \\
000 & 3 & [s1, s3] \\
001 & 3 & [s1, s4, s6] \\
010 & 3 & [s1, s4, s7, start, s1] \\
011 & 3 & [s1, s4, s7, start, s2] \\
100 & 3 & [s2, s4] \\
101 & 3 & [s2, s5, s7] \\
110 & 3 & [s2, s5, s8, start, s1] \\
111 & 3 & [s2, s5, s8, start, s2] \\
0000 & 4 & [s1, s3, start, s1] \\
0001 & 4 & [s1, s3, start, s2] \\
0010 & 4 & [s1, s4, s6, start, s1] \\
0011 & 4 & [s1, s4, s6, start, s2] \\
0100 & 4 & [s1, s4, s7, start, s1, s3] \\
0101 & 4 & [s1, s4, s7, start, s1, s4] \\
0110 & 4 & [s1, s4, s7, start, s2, s4] \\
0111 & 4 & [s1, s4, s7, start, s2, s5] \\
1000 & 4 & [s2, s4, s6] \\
1001 & 4 & [s2, s4, s7] \\
1010 & 4 & [s2, s5, s7] \\
1011 & 4 & [s2, s5, s8, start, s1, s3] \\
1100 & 4 & [s2, s5, s8, start, s1, s4, s6] \\
1101 & 4 & [s2, s5, s8, start, s1, s4, s7] \\
1110 & 4 & [s2, s5, s8, start, s2, s4] \\
1111 & 4 & [s2, s5, s8, start, s2, s5] \\
00000 & 5 & [s1, s3, start, s1, s3] \\
11111 & 5 & [s2, s5, s8, start, s2, s5, s8, start, s2] \\
	\hline
\end{tabular}
\caption{Table of example encodings using $encode_{fixed}$. Two different inputs can encode to the same text only if they have a different length, as happens with $00$ and $000$. The frequencies of the different bigrams approximate the probabilities $P(s_{n} | s_{n - 1})$ from the Markov chain, and this approximation gets better as $n$ grows (because the space is bigger, and because of the Condition of Minimal Length, which has a higher effect in smaller inputs). Also this table shows a very low capacity, because the Markov chain used is very small. More comments about capacity in Section~\ref{conclusion}.}
\label{fig:tableencodings}
\end{table}

\subsection{Decoding of Fixed-size Data Using Markov Chains} \label{decodingfixed}

Decoding of fixed-size data is based on $subrangeForState$, which was described on Section~\ref{mappingprob}. It was previously described as a function that returns the subrange that is assigned to a given state; but it can also be seen as a decoder from states to numbers. In this way, the function $subrangeForState(mc, w_{k}, r, w_{l})$ decodes a single word state $w_{l}$, given that the previous state was $w_{k}$. The decoded value is not a number, but a range of numbers: $[a, b]$ where both $a$ and $b$ are natural numbers.

Given an input sequence of states or words $w_{t}$ (where $w_{0}$ is taken to be the initial state used for encoding) and an initial range $r_{0}$ (typically $[0, 2^n - 1]$) we define the sequence of ranges $r_{t}$ as:

\begin{equation}
r_{t} = subrangeForState(mc, w_{t}, r_{t - 1}, w_{t - 1})
\end{equation}

The output of the $decode_{fixed}$ is the value of the range $r_{T}$, where $T$ is the first $t$ such that $length(r_{t}) = 1$. Since when that happens the range covers a single number, the decoding process can just return that number.

A valid output isn't guaranteed for all texts (sequences of words), only for words that have been generated by using the $encode_{fixed}$ process described above.

The decoding process works because it follows the same path that the encoder process followed when generating the text, and this path leads to the original input data. The encoder writes a sequence of words while refining subranges until finding a range that has length $1$. The decoding process follows the states written by the encoder, which lead to exactly the same sequence of subranges. This means that $decode_{fixed}$ will reach the input of $encode_{fixed}$, when feed with the output of $encode_{fixed}$. This makes $decode_{fixed}$ acts as the inverse for $encode_{fixed}$, for fixed $n$.

An additional property of $decode_{fixed}$ as it is defined here is that if $data = decode_{fixed}(text)$, then also $data = decode_{fixed}(text + text_{2})$, where $text_{2}$ is any text and "$+$" is the list concatenation operation. This is because the fixed decoding algorithm finishes computing the value for $data$ when the last subranges converge to a single number, and that happens at the same place in the text sequence for $text$ and for $text + text_{2}$.

This property is useful because it allows us to concatenate encoded texts, and they can be decoded directly as the decoder can tell where every text starts and ends. This is applied to the variable encoding algorithm discussed in Section~\ref{variableencoding}.

\subsection{Implementation Details}

A direct implementation of the algorithms described above would require that many operations are applied to the $n$ bit ranges in every iteration of encoding and decoding. For example, in every iteration of the fixed-size decoding algorithm, a call to $subranges$ needs to be done with a range of numbers with $n$ bits of size, until the length of the selected range is $1$ (so that the range matches the original input). This is very inefficient both regarding memory usage and processing time.

It is possible to avoid processing on the full $n$ bits on every iteration, by making some changes to the underlying algorithms. Some data with length $n$ can be processed more efficiently if only a short, moving window of a few bits is processed in every iteration. We define $subranges_{fast}$:

\begin{equation}
subranges_{fast}(mc, s, r_{m bits}, n) = subranges(mc, s, expand(r_{short}, n))
\end{equation}

where $r_{short} = [a, b]$ is defined to be a range where $a$ and $b$ are two numbers that can be expressed in up to $m$ bits, and $expand(range, m, n)$ computes $[a_{2}, b_{2}]$, with $a_{2}$ identical to $a$ in all its leftmost $m$ bits, and $0$ in the remaining bits, and with $b_{2}$ identical to $b$ in all its leftmost $m$ bits, and $1$ in the remaining bits. This means that we can use $expand$ to convert short ranges like $[01, 10]$ (in binary) to the longer 4 bit range $[0100, 1011]$, if $n = 4$.

An efficient implementation of $subranges_{fast}$ returns all subranges in short form, for any input. When the ranges have to be split in a way that requires infinite or long precision (for example if there are two states, with $P(s_{1} | s) = 0.3$ and $P(s_{2} | s) = 0.7$), this is only possible if a precision limit is set in the implementation. This precision limit can be set to mean that regardless of the input of $subranges_{fast}$, there is a maximum number of bits that can be used for the partitioning process.

For example, with $n = 100$ and the probabilities described above, the ranges returned could be: [00000000, 01001101] for $s_{1}$, and [01001110, 11111111] for $s_{2}$. In this case, $s_{1}$ really has about 0.305 of the numbers of the total range, so using 8 of the 100 bits is a good approximation. If we were to use only 4 bits in $subranges_{fast}$ for this case, it would return: [0000, 0100] for $s_{1}$, and [0101, 1111] for $s_{2}$. In this case $s_{1}$ maps to about 0.312 numbers of the total range; this is a slightly worse approximation, but it might be better as it requires using only half the amount of bits.

Both for encoding and decoding, a bit stream data structure will be needed. For encoding, this stream of bits will be read; when decoding, it will be used to write the data output, in a bitwise fashion.

When encoding, in every iteration $subranges_{fast}$ will require a small number of bits to be read from the bit stream. As soon as those bits are read, they can be discarded from the bit stream. Also, $subranges_{fast}$ will generate new ranges in every call, and in every iteration these ranges will be more precise, that is, ranges that cover a smaller amount of numbers. This means that the subranges will require more bits to be stored.

However, if the precision for $subranges_{fast}$ is set to a finite value (as described above), the number of bits at the right of the range that differ from each other will be at most $k$, for some $k$. This means that with every iteration, the ranges will grow in size $n$, but the leftmost bits will at the same time converge bitwise to the same values (for range $[a, b]$, leftmost bits of $a$ and $b$ will be identical). The leftmost bits can then be discarded, as they are already known to match the leftmost bits in the input data.

This process ensures that in every iteration of encoding, $subranges_{fast}$ only has to deal with a moving window that has a limited number of bits, related to the precision set to the system in the implementation.

Similarly for decoding; in very iteration, the range that $subranges_{fast}$ returns will grow in size (as measured in bits). However, while the range grows in size, the leftmost bits converge, so they can be removed, and added to an output bit stream. When the process finishes, the output bit stream will contain the full output of the decoding algorithm: all the bits of the converged range.

\begin{figure}

\begin{tabular}{|c|c|c|c|c|c|}
	\hline
\rule{0pt}{0.45cm} \rule[-0.30cm]{0pt}{0pt} image & chain states & encoding time & decoding time & encoded size & $\frac{\mbox{encoded size}}{\mbox{file size}}$ \\
	\hline
example.zip & unigrams & 91.8 s & 95.3 s & 81 kB & 6.7  \\
(12 kB) & & \emph{(0.1 kB/s)} & \emph{(0.1 kB/s)} & \emph{(zip: 32 kB)} & \emph{(zip: 2.7)}  \\
\cline{2-6}
 & bigrams & 53.7 s & 55.8 s & 149 kB & 12.4 \\
 & & \emph{(0.2 kB/s)} & \emph{(0.2 kB/s)} & \emph{(zip: 58 kB)} & \emph{(zip: 4.8)} \\
	\hline
example.jpg & unigrams & 141 s & 150.5 s & 119 kB & 6.3 \\
(19 kB) &  & \emph{(0.1 kB/s)} & \emph{(0.1 kB/s)} & \emph{(zip: 38 kB)} & \emph{(zip: 3.2)} \\
\cline{2-6}
(zip: 12 kB) & bigrams  & 93.4 s & 97.6 s & 216 kB & 11.4 \\
 &  & \emph{(0.2 kB/s)} & \emph{(0.2 kB/s)} & \emph{(zip: 66 kB)} & \emph{(zip: 5.5)} \\
	\hline
example.png & unigrams & 299.6 s & 317.2 s & 269 kB & 6.9 \\
(39 kB) &  & \emph{(0.1 kB/s)} & \emph{(0.1 kB/s)} & \emph{(zip: 104 kB)} & \emph{(zip: 2.7)} \\
\cline{2-6}
 & bigrams  & 194.1 s & 181.8 s & 494 kB & 12.7 \\
 &  & \emph{(0.2 kB/s)} & \emph{(0.2 kB/s)} & \emph{(zip: 188 kB)} & \emph{(zip: 4.8)} \\
	\hline
\end{tabular}
\vspace{1em}

\small{All benchmarks were run on a computer with processor Intel Core i7-2670QM CPU at 2.20GHz x 8, with 7.7 GiB RAM.}

\vspace{1em}

\framebox{
\parbox{12cm}{
Samples of the encoded texts:

\begin{itemize}

\item Example.zip (unigrams): "Be limited and secondly because Pierre suddenly realized. Und die and secondly. Monotonous sound of the man who too late. Monsieur Kiril Andreevich nicknamed the hour later grasped the Russian commanders. Dressed for the new building with a year period of the two or an example."

\item Example.zip (bigrams): "Be a square for fuel and kindled fires there. Secondly it was hard to hide behind the cart and remained silent. He feels a pain in the now cold face appeared that the man continually glanced at her as though they stumbled and panted with fatigue. With a deep."

\item Example.jpg (unigrams): "He had been her neighbors and friends that these wrinkles and then there's no lambskin cap and saw that Russian expedition. Under a largish piece of me all all is going on the lot of that moment I have an all four abreast. Having evidently relating to scrutinize the nunnery. Every moment."

\item Example.jpg (bigrams): "He had something on both sides and. Secondly it was tete a tete. You did me the duty of a month ago. He's having a good humored amiable smiles. Pierre pointed to a series of actions that follows therefrom. After playing out a passage she had all the forms of town life perished. Tell him Here."

\item Example.png (unigrams): "Rostov a short fingers and exhausted and the driver a bright lilac dress. And the locomotive by all seemed to the most profitable source of Karataev and secondly. Even remember that it an enormous movements and had to tell you want of the other troops standing. If it's high."

\item Example.png (bigrams): "Rostov looked inimically at Pierre and addressing all present and rested on them. But seeing before him. Princess Mary thought only of how Princess Mary for Prince Vasili saw that Platon did not forget what I consider myself bound to Princess Mary will take the covert at once abandoned all their decorations."

\end{itemize}
}}

\vspace{0.5em}
\caption{Example benchmarks and results when running MarkovTextStego \cite{moraldo} with Markov chains generated from War and Peace by Tolstoy. MarkovTextStego uses the method discussed in this article, and an extension that uses bigrams as states in the Markov chain (as discussed briefly in Section~\ref{markovmodels}). The bigram-based encoder will produce higher quality texts, however they will be larger than those produced by the unigram-based encoder.}
\label{fig:tablefiles}
\end{figure}

\section{Variable Size Encoding and Decoding} \label{variableencoding}

The encoding and decoding process described above only allows to decode data from a text, given that the size of the data is known beforehand. However, requiring the recipient of a steganographic system to know the size of the hidden data before it is decoded is not optimal. An extension of the encoding and decoding methods for variable-size data solves this problem.

For variable size encoding and decoding it is required that an integer $m$ is shared beforehand. This number is not the data size, but the size used for a header; it is typically a small value like 16 or 32. Texts $c_{1}$ and $c_{2}$ are encoded as shown below, using the three arguments version of $encode_{fixed}$.

The header is encoded first, into $c_{1}$. This is done using the fixed-data encoding algorithm, with the fixed size $m$ that is known both for encoder and decoder:

\begin{equation}
n = length(data)
\end{equation}

\begin{equation}
c_{1} = encode_{fixed}(n, m, start)
\end{equation}

Once the header was encoded, the actual data is encoded into $c_{2}$. We use $w$ as starting symbol, to ensure that there isn't an interruption in the flow of the generated text between the last symbol in $c_{1}$ and the first one in $c_{2}$:

\begin{equation}
w = \mbox{ last word in } c_{1} \mbox{ text sequence}
\end{equation}

\begin{equation}
c_{2} = encode_{fixed}(data, n, w)
\end{equation}

Finally, $encode(data)$ is defined simply as:

\begin{equation}
encode(data) = c_{1} + c_{2}
\end{equation}

That is, the encoded data is just the header text followed by the data text. As $w$ was used as starting symbol for generating $c_{2}$, there will be no interruption in the flow between both texts.

For decoding an input $text$, we define:

\begin{equation}
n\prime = decode_{fixed}(text, m, start)
\end{equation}

That is, $decode_{fixed}$ is used to extract the length information from the header, using the shared value $m$.

\begin{equation}
text_{1} = \mbox{ list of words used in decoding } n\prime
\end{equation}

\begin{equation}
text_{2} = \mbox{ list of words not used in decoding } n\prime
\end{equation}

\begin{equation}
w\prime = \mbox{ last of } text_{1}
\end{equation}

Finally, $decode$ can be defined:

\begin{equation}
decode(text) = decode_{fixed}(text_{2}, n\prime, w\prime)
\end{equation}

It can be seen that when $data\prime = decode(encode(data))$, it follows that: $n\prime = n$, $text_{1} = c_{1}$, $text_{2} = c_{2}$, and $w\prime = w$. For this reason, $data\prime = data$, which means that $decode$ is the right decoding function.

It is also possible to extend $encode$, without changing this last property, in this way:

\begin{equation}
encode(data) = c_{1} + c_{2} + randomText(z)
\end{equation}

where $z$ is the last word in $c_{2}$, and $randomText(symbol)$ generates a random text that ends in period, using the Markov chain and starting from the given state. This can be used to ensure that all texts generated by $encode$ have a final sentence that is complete, and finishes with period. Adding any text won't affect the decoding at all, as explained in Section~\ref{decodingfixed}.

Depending on the kind of data that is being transmitted, it might be useful to encode into $c_{1}$ the length of the data in bytes, instead of encoding it in bits. Also, the way the length is actually represented into bits matters; if big endian is used to represent a multi-byte length into bytes, short encoded lengths will start with a sequence of 0 bits; this could produce the encoded texts to always start with the same words, or with a small variety of different words (because all leftmost bits are zero). For this reason, either little endian or a representation that reverses the bits of big endian would be preferable.

\section{Conclusions and Future Research}\label{conclusion}

This article presented a steganographic method based on Markov chains that differs from other similar models in the way precision loss in the language model is avoided. A reference implementation for this method was also presented.

The examples shown in Table~\ref{fig:tableencodings} could seem to show that the system has very low capacity. However this is only because of the Markov chain used; if the system uses a small Markov chain, it will have low capacity, but if it uses a bigger Markov chain it will typically have a higher capacity.

Preliminary results of empirical tests using a big Markov chain computed from an actual literary text show that the encoded data takes the size of about 6 - 7 times the size of the original data, with an $n$ value that is big enough (for very small $n$, yet bigger than a few bytes, this factor can be higher, e.g. around 9). Because the produced output is a text, it can be compressed with a high ratio; the compressed size of the texts is about 2 times the size of the original data. However, these results require a more complete and thorough analysis.

Other possibilities for further research are: to combine this method to other known language based steganographic systems, for producing an overall better steganographic text generation method; to analyze what is the actual, measured performance for this new algorithm, and how this new algorithm compares to other existing algorithms, in terms of stegoanalysis.

\end{document}